\documentclass{optica-article}

\journal{opticajournal} 

\articletype{Research Article}

\usepackage{lineno}
\usepackage{comment}
\usepackage{graphicx}    
\usepackage{hyperref}
\usepackage{textcomp}
\usepackage{ulem}
\usepackage{helvet}
\usepackage{tikz}
\newcommand*\circled[1]{\tikz[baseline=(char.base)]{
            \node[shape=circle,draw,inner sep=0.4pt] (char) {#1};}}

\begin{document}

\title{Double-pass rotating z-cut quartz plate as a rapidly variable waveplate}

\author{Byungjin Lee,\authormark{1} Kiryang Kwon,\authormark{1} and Jae-yoon Choi\authormark{1,$\dagger$}}

\address{\authormark{1}Department of Physics, KAIST, Daejeon 34141, Korea}
\email{\authormark{$\dagger$}jaeyoon.choi@kaist.ac.kr} 

\begin{abstract*} 
We demonstrate a rapidly tunable waveplate based on a rotating z-cut quartz plate in a double-pass configuration. 
In contrast to previous single-pass implementations, where angular rotation of birefringent crystals causes significant beam path displacement, we show that the double-pass geometry effectively suppresses beam walk-off, reducing lateral shifts to below 10~$\mu{}$m, which is stable enough to have a fiber coupling. 
We present a full theoretical description of the polarization changes using Jones matrix calculations and verify it through polarization-resolved measurements. 
Additionally, the retardation is stable across a broad spectral range without requiring wavelength-specific optimization. 
When combined with a polarizing beam splitter, the system functions as a high-speed optical power modulator, achieving a dynamic power conversion in 1~ms with its contrast about 1000:1.
This compact and robust design is particularly suited for atomic, molecular, and optical (AMO) physics experiments requiring rapid and precise control of light intensity.
\end{abstract*}

\section{Introduction}
The ability to manipulate the polarization state of light is essential for many optical applications, such as quantum optics~\cite{Aspect1982}, spectroscopy~\cite{Pearman2002}, nonlinear optics~\cite{Franken1961}, microscopy~\cite{Arteaga2009} and biomedical applications~\cite{He2021}.
Birefringent crystals are widely employed in such tasks due to their ability to induce a controlled phase retardation between orthogonal polarization components. Among these materials, quartz has been extensively used, especially in terahertz (THz) optics, owing to its high transparency~\cite{Kaveev2014,Davies2018,Zhang2023}, mechanical and thermal stability~\cite{Gotze2009}, and high laser damage threshold~\cite{Balos2023}. The ability to manipulate the effective retardation by varying the angle of incidence forms the basis of the Berek compensator~\cite{Fisher1948}, a classic optical device still used in modern polarization-sensitive applications such as THz detection~\cite{Frenzel2024} and polarized microscopy~\cite{Shribak2015}.

Recently, the rotation of z-cut uniaxial quartz plates has been adopted to control laser power in atomic, molecular, and optical (AMO) physics experiments. 
Compared to conventional Pockels cells based on the electro-optic effect, quartz offers superior thermal stability and high-power compatibility. 
Its thermal expansion coefficient is one to two orders of magnitude smaller than that of typical EOM crystals, mitigating thermal lensing and optical degradation.
Leveraging these advantages, the rotating z-cut quartz method has been applied to quantum gas microscope setups,  enabling high-speed beam power manipulation~\cite{Blatt2015,Mazurenko2019} and phase control~\cite{Li2021}. 
However, a major limitation of the single-pass configuration is the beam path shift caused by the rotation of the birefringence plate, which introduces challenges for alignment-sensitive applications or systems requiring high beam-pointing stability.

In this work, we present a compact and practical solution to this limitation, based on a double-pass configuration of a rotating z-cut quartz plate [Fig.~1(a) and (b)]. 
This setup significantly mitigates beam path displacement by reflecting the beam back through the same optical path as the incident beam, reducing lateral shifts to be compatible with fiber coupling. 
We also provide a theoretical background, deriving the full Jones matrix formalism for both single- and double-pass geometries. 
We experimentally validate polarization transformation across a broad wavelength range. 
Notably, the device operates as a variable waveplate capable of achieving full polarization rotation with only $\sim0.7^{\circ}$  angular modulation, in contrast to typical zero-order waveplates that require a $45^{\circ}$ rotation.
This small angular requirement enables sub-millisecond switching times.
Lastly, we show the angular vibrations from the rapid motion can also be suppressed, where the light intensity fluctuations can be below 0.1$\%$. 
In Sections 2 and 3, we investigate the polarization control and beam path stability in detail. 
In Section 4, we demonstrate the implementation of this device as a rapid optical power modulator, achieving a dynamic range of 1000:1 within 1 ms. \\

\begin{figure}[htbp]
\centering\includegraphics[width=0.9\linewidth]{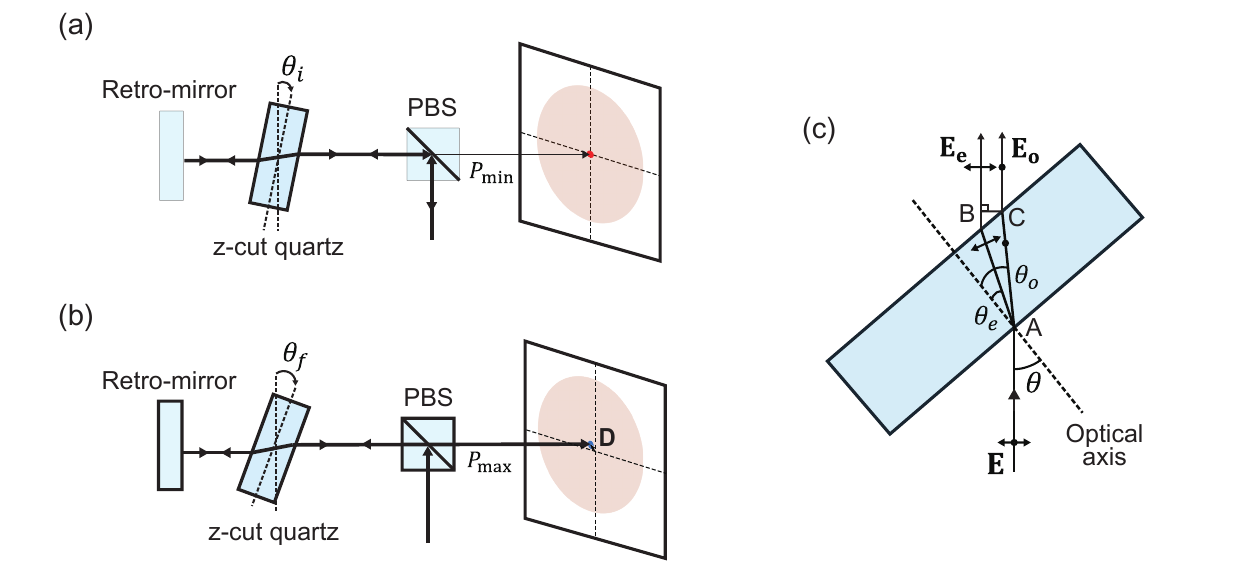}\label{Fig1}
\caption{Z-cut quartz plate in the double pass configuration is illustrated in (a) and (b). The conversion between (a) and (b) is performed within a 1 ms timescale, resulting in a beam power variation across a dynamic range of $1000:1(P_{\mathrm{max}}:P_{\mathrm{min}})$ in our setup. The beam path shift $D$ is under $10 \,\mathrm{\mu m}$, largely reduced compared to the single-pass configuration. (c) Illustration of beam propagation as a function of quartz angle $\theta$. The incident electric field $\mathbf{E}$ refracts with different angles $\theta_o$ and $\theta_e$ because of the birefringence, having a finite displacement between the extraordinary $\mathbf{E_e}$ and ordinary $\mathbf{E_o}$ light after passing through the quartz. 
}
\end{figure}

\section{Polarization control by the rotating z-cut quartz}\label{Sec2}
The working principle of polarization control by rotating the z-cut quartz is summarized in Fig.~1(c)~\cite{Durey2021}. 
Because of the birefringence of the quartz, the ordinary ($\mathbf{E}_{\rm o}$) and extraordinary ($\mathbf{E}_{\rm e}$) components undergo a different beam path and experience a phase retardation. 
The beam path difference can be calculated from Snell's law, so that the tilting angle $\theta$ of the quartz determines the polarization of the light. 
Although these two beams are separated by a finite distance, denoted as $BC$ in \hyperref[Fig1]{Fig.1(c)}, the distance is much smaller than the typical beam waist~\cite{Holmes1964,Ghosh1999}. 
For example, when a quartz has 10 mm thickness and is rotated by $20^\circ$, the beam separation is only $13.2 \,\mathrm{\mu m}$ for 1064 nm light.

\subsection{Theory for rotating z-cut quartz}
Assuming a collimated and monochromatic laser beam, the analytic expression of the phase retadation $\Delta \phi(\theta)$ as a function of rotation angle $\theta$ can be determined from the path difference between the ordinary and extraordinary beam~\cite{Durey2021}, which is written in the following equation, 
\begin{equation}
    \Delta\phi(\theta) = 2\pi \frac{\Gamma(\theta)}{\lambda} =\frac{2\pi}{\lambda}n_od\left (\sqrt{1-\frac{\sin^2{\theta}}{n_e^2}}-\sqrt{1-\frac{\sin^2{\theta}}{n_o^2}}\right).
\end{equation}
Here, $\lambda$ is the wavelength of the beam, $d$ is the width of the z-cut quartz plate, and $n_o, n_e$ are refractive indices for ordinary and extraordinary components, respectively, and the optical activity is not considered (\textit{Appendix II}).  
Quartz is a positive birefringence medium($n_o<n_e$), so the ordinary component is a fast axis and the extraordinary component is a slow axis. For negative  birefringence medium, the sign of the $\Delta \phi$ changes.

In our experiments, the rotating axis of the quartz becomes the fast axis, and $\beta$ is defined as an angle between the horizontal $x$-axis and the fast axis [Fig.~2(a)]. 
Then, the Jones matrix of rotating z-cut quartz $\mathbf{Q}(\beta,\theta)$ can be expressed as,  
\begin{equation}
    \mathbf{Q}(\beta,\theta)=\begin{pmatrix}
Q_{xx} &Q_{xy}\\
Q_{yx}&Q_{yy} \end{pmatrix} =  \begin{pmatrix}
\cos^2{\beta}+e^{i\Delta\phi}\sin^2{\beta} &(1-e^{i\Delta \phi})\sin{\beta}\cos{\beta}\\
(1-e^{i\Delta \phi})\sin{\beta}\cos{\beta}&\sin^2{\beta}+e^{i\Delta\phi}\cos^2{\beta} \end{pmatrix}.
\end{equation}
If quartz rotating axis is parallel or vertical to $x$-axis, $\beta=0,\,\pi/2$. 
Then, P- or S-polarization beams do not experience polarization rotation due to $\mathbf{Q}(\beta,\theta)$ becoming a diagonal matrix, and it can't be used as a retarder. 
To have the off-diagonal components, the angle is set to $\beta=\pi/4$ by mounting the rotating motor with $45^\circ$ tilted from the ground, Fig.~2(b). 
For the retro-reflected beam in the double pass configuration, where the light propagates along the $-z$-axis, the angular variables change their sign to have $(-\beta,-\theta)$. 
Since the $\Delta\phi(\theta)$ is symmetric for quartz angle $\theta$, the Jones matrix representation for the reflected beam is $\mathbf{Q}(-\beta,\theta)$ [Fig.~2(c)].

\begin{figure}[htbp]
\centering\includegraphics[width=0.9\linewidth]{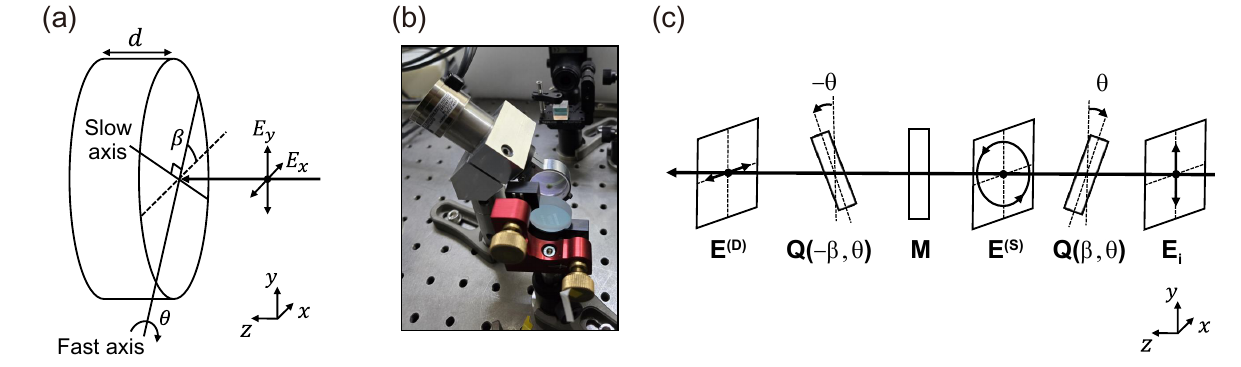}\label{Fig2}
\caption{Layout of the configuration when the beam is incident on a z-cut quartz plate. (a) For the z-cut quartz, the rotating axis corresponds to the ordinary component, and the extraordinary component is vertical. Z-cut quartz is a uniaxial crystal, where the optical axis of the quartz is aligned along the $z$-direction in this figure. The slow and fast axes remain unchanged, even when the z-cut quartz is rotated around the $z$-axis. $\beta$ is an angle between $x$-axis and fast axis, and $\theta$ is a quartz angle. 
(b) Image of the optics system. Quartz attached motor (Galvanometer GVM-2510, Citizen Chiba) is fixed on a $45^\circ$ tilted aluminium mount.
(c) Changing the polarization of the incident S-polarization beam by rotating the quartz angle $\theta$.
Electric fields $\mathbf{E^{(S)}}$($\mathbf{E^{(D)}}$) for the single (double) pass configuration can be calculated from the Jones matrix. Illustrated polarization is for the $\Delta\phi=\pi/2$ case.}
\end{figure}

Using the Jones matrix, we obtain the electric field of the incident beam for the single-pass $\mathbf{E^{(S)}}$ and double-pass $\mathbf{E^{(D)}}$ configuration,
\begin{equation}
    \mathbf{E^{(S)}}=
    \begin{pmatrix}
    E^{(S)}_x\\
    E^{(S)}_y
    \end{pmatrix}=
    \mathbf{Q}(\pi/4 ,\theta) \cdot \mathbf{E_i}=
    \frac{1}{2}
    \begin{pmatrix}
    1-e^{i\Delta\phi}\\
    1+e^{i\Delta \phi} 
    \end{pmatrix},
\end{equation}
\begin{equation}
\begin{split}
    \mathbf{E^{(D)}}&=
    \begin{pmatrix}
    E^{(D)}_x\\
    E^{(D)}_y
    \end{pmatrix}
    =\mathbf{Q}(-\pi/4 ,\theta)\cdot\mathbf{M}\cdot\mathbf{E^{(S)}}
   =\frac{1}{2}
    \begin{pmatrix}
    1-e^{i2\Delta\phi}\\
    -(1+e^{i2\Delta \phi}) 
    \end{pmatrix},
\end{split}
\end{equation}
with $\mathbf{M}$ denotes the Jones matrix of a mirror under normal incidence. Therefore, changing the rotation angle $\theta$, we can vary the retardation $\Delta\phi(\theta)$ and the polarization of the incident light.
Together with the polarizing beam splitter (PBS) cube, the polarization changes are converted into a variation of the laser power. In this example, the transmitted laser beam power after PBS for single-pass $P_x^{(S)}$  and double-pass $P_x^{(D)}$ is given by,
\begin{equation}
    P_x^{(S)}=|E^{(S)}_x|^2=\sin^2{(\Delta \phi/2)}, 
\end{equation}
\begin{equation}
    P_x^{(D)}=|E^{(D)}_x|^2=\sin^2{(\Delta \phi)}.
\end{equation}

\begin{figure}[t]
\centering\includegraphics[width=\linewidth]{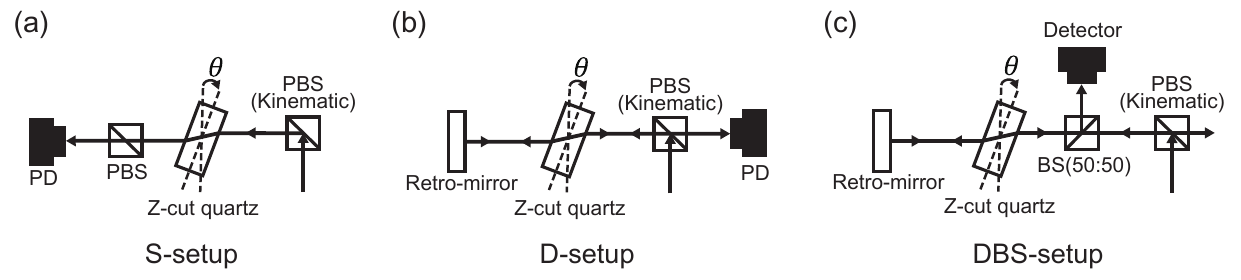}\label{Fig3}
\caption{(a) S-setup for single-pass configuration. (b) D-setup for double-pass configuration. Polarization-dependent power can be measured using the S-setup and the D-setup. (c) DBS setup for double-pass configuration. Beam pointing can be measured directly using the DBS setup.}
\end{figure}

\begin{figure}[b]
\centering\includegraphics[width=\linewidth]{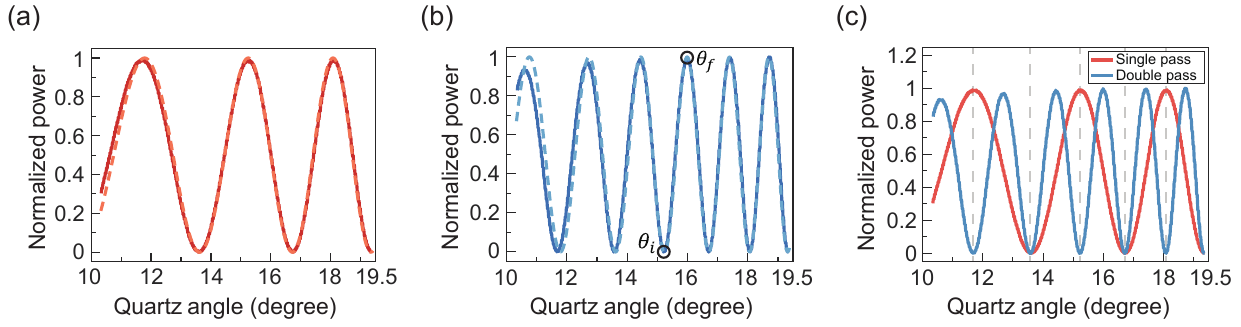}\label{Fig4}
\caption{Normalized measured beam powers for the (a) single-pass and (b) double-pass cases. The dashed lines in (a) and (b) represent the theoretical curves, while the solid lines indicate the measured power curves. For the power conversion application, two angles, such as $\theta_i$ and $\theta_f$, are used to form the $\pi/2$ retardation peak. In our case, the power conversion angle is $\theta_f-\theta_i=0.7^\circ$.  (c) shows the overlapped graph of the two measured power curves from (a) and (b), demonstrating that the period of the double-pass power curve is half that of the single-pass curve. Gray dashed lines represent the minimum angle of the double-pass configuration. Measured power curves are obtained by averaging 10 measurements, and we use 2.8~mW laser beam at 1064~nm wavelength.}
\end{figure}

\subsection{Polarization control by rotating quartz angle}
To verify these theoretical predictions, we measure the power transmission of the single- and double-passed beams through a PBS as a function of quartz angle (Fig.~3).
We use 1064 nm wavelength continuous-wave laser light  with beam diameter of $1/e^2=2.2 \,\mathrm{mm}$ and divergence of approximately $100\,\mathrm{\mu rad}$. 
The quartz plate (Cening optics, AR coated at 1064 nm) has $d=10\, \mathrm{mm}$ width, where refractive indices are $n_o=1.534$ and $n_e=1.543$ at 1064~nm~\cite{Ghosh1999}.  
The galvanometer (GVM-2510, Citizen Chiba) can rotate the quartz with a repeatability of 8 $\mu$rad. 
In the experiment, we rotate the quartz angle from $10^\circ$ to $20^\circ$ to minimize the effect of the reflected beam from the crystal surface~\cite{Pietraszkiewicz1995}. 
Even within this range, it is sufficient to achieve full control of polarization and transmitted laser beam power.
Figure 4 displays the measured power curves with different angle $\theta$ both in the single and double pass configurations. 
The experimental result shows excellent agreement with the theoretical predictions.
Moreover, as expected from Eq.~(5) and (6), we find that the double-pass power curve has half the period of the single-pass. 
It can be shown by overlapping the two results in Fig.~4(c), where the minimum (maximum) angle of the single pass configuration corresponds to the even (odd) minimum angle of the double pass setup. 
This feature highlights that the double-pass configuration has the advantage of rapidly controlling polarization over the single-pass setup.

\subsection{Rotating z-cut quartz as a waveplate available in a wide wavelength range  }
 As indicated in Eq.(1), rotating z-cut quartz is able to change the beam polarization across various wavelengths with corresponding $\Delta\phi (\theta)$ [Fig.~5(a)]. 
To verify this in an actual experiment, we direct a 671 nm beam onto a quartz plate and measure the power using S- and D-setups. The measured result is shown in  Fig.~5(b), and the power contrast varies similarly to the 1064 nm case. Due to the retardation for 671 nm is larger than 1064 nm at the same quartz angle, there are more power peaks in the same angle range compared to Fig.~4(c). Furthermore, the period of double-pass configuration is half of the single-pass, so Eq.(5) and Eq.(6) hold well. In Fig.~5(b), the power contrast of the double-passed curve increases with the quartz angle. This could be attributed to the anti-reflection coating of the quartz being optimized for 1064 nm rather than 671 nm.  Therefore, z-cut quartz plate could be used for a wide range of single-frequency laser light with a linewidth order of 1~MHz or less.

\begin{figure}[htbp]
\centering\includegraphics[width=\linewidth]{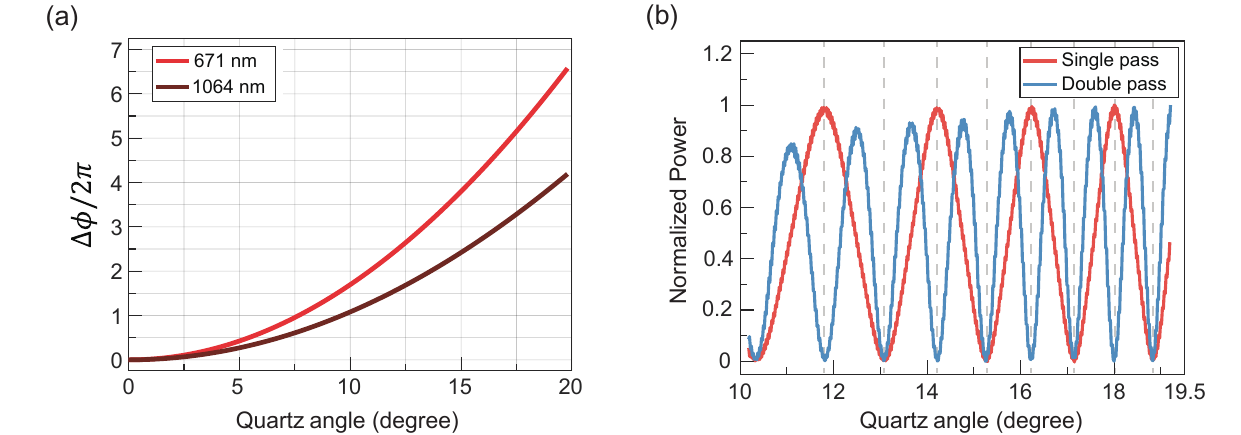}\label{Fig5}
\caption{(a) Retardation $\Delta \phi$ from Eq.(1) for 671 and 1064 nm. Wavelength term $\lambda$ is in the denominator, so retardation for 671~nm is larger than 1064~nm. (b) Measured power for single and double-pass configurations with a 671 nm beam. Measured power contrast is varied well, and the period of the double-pass configuration is half of the single-pass configuration as expected. Vertical dashed lines denote the minimum angle of the double-pass configuration. Measured power curves are obtained by averaging 10 measurements and the input laser beam power is 2.5~mW.}
\end{figure}

\section{Beam path shift for single- and double-pass configuration}\label{Sec3}
One of the main limitations of using the rotating quartz as a variable retarder is the beam point shift caused by the quartz rotation~\cite{Mazurenko2019}. 
In this section, we present how the double-pass configuration can overcome this difficulty. 
The shift of the incident beam can be understood from simple geometrical optics, Fig.~6(a). 
The beam path shift in a single pass configuration, for example, for the ordinary (extraordinary) beam can be expressed as,
\begin{equation}
    D_o=\frac{d}{\cos{\theta_o}}\sin{(\theta-\theta_o)},
\end{equation}
\begin{equation}
    D_e=\frac{d}{\cos{\theta_e}}\sin{(\theta-\theta_e)}.
\end{equation}
\noindent Here, $\theta_o$ and $\theta_e$ are the angles of refractive from Snell's law: $n_e\sin{\theta_e}=n\sin{\theta}=n_o\sin{\theta_o}$ with air refractive index $n$. 
Since the z-cut quartz has a small refractive index difference of $\Delta n = n_e-n_o=0.009$ for a 1064 nm beam, the displacements $D_o$ and $D_e$ are nearly identical, Fig.~6(b).

In the experiment, we can directly monitor the angle dependence of the beam center by putting a 2D lateral sensor (PDP90A, Thorlabs) at the position of the PD in the S-setup. 
The incident beam has a linear polarization along the vertical $y$-axis, having both ordinary and extraordinary components. 
The measured beam path shift $D_M$ increases with the rotation angle [Fig.~6(b)] and shows a good agreement with the simple theory curve.

\begin{figure}[htbp]
\centering\includegraphics[width=0.9\linewidth]{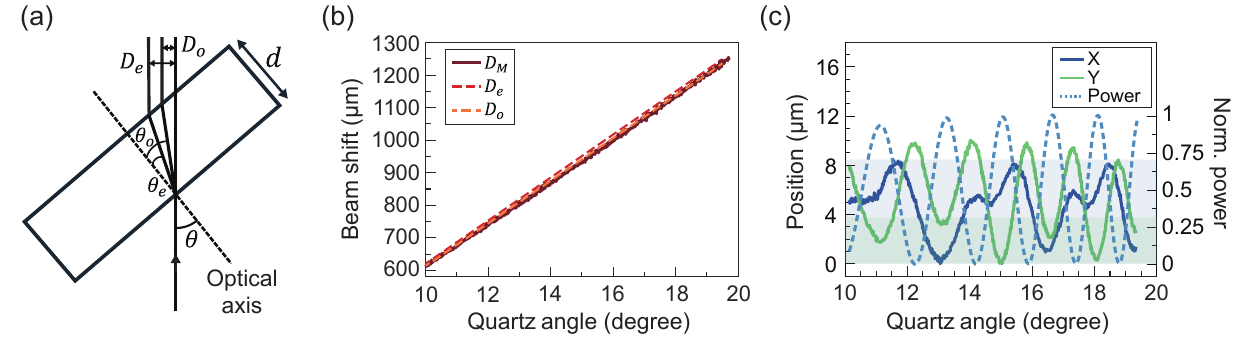}\label{Fig6}
\caption{
(a) Layout of incident beam and quartz plate for geometrical optics derivation. (b) Beam path shift of single-pass configuration with theoretical curves. The reference for the position shift is the beam position when the quartz angle is zero.  $D_M$ closely follows the trend predicted theoretically. $D_M$ is obtained by averaging 10 measurements. (c) Power curve and X (along $x$-axis) and Y (along $y$-axis) position variance of double-passed beam with respect to quartz angle. Because of its rapid rotation speed, there is an angle offset of about $0.7^\circ$ compared to Fig.~4.
The shading represents the beam shift caused by the ordinary HWP, measured as $(8.5, 3.8)\, \mathrm{\mu m}$ for the $x$ and $y$ axes. The measured X and Y values are obtained by averaging 20 measurements and {applying a moving average filter~\cite{Chang2011,Chung2015,Li2023} with a window size of 5 to reduce background electronic noise.}
}
\end{figure}

In the double-pass configuration, no position shift is expected because the retro-reflected beam follows the incident beam path.
We experimentally study the beam position shift by picking up the retro-reflected beam with a beam splitter (BS) [DBS-setup in Fig.~3(c)]. 
This is because optimal beam power range of the position sensor is narrow, we cannot directly measure the beam shift during power conversion using the D-setup.
The result is displayed in Fig.6(c), where we observe oscillation of the beam position $X$ and $Y$ as a function of the quartz rotation angle. 
In this measurement, the quartz-mounted motor rotates by $10^\circ$ within a 20~ms period to mitigate low-frequency beam point fluctuations in our optical system. 
This oscillatory behavior is attributed to the property of the BS that has a weak position dependence of the incident polarization. 
To test the polarization dependence position shift of the BS, we create S- and P-polarizations by rotating a zero-order half-waveplate (HWP) and measure the beam position of the reflected beam by the BS.
The beam shift caused by the HWP is measured to be $(8.5, 3.8) \,\mathrm{\mu m}$ for the $x$ and $y$ axes, which is comparable to  the oscillation amplitude measured with rotating quartz. 
The polarization dependence on the beam splitter also explains that the beam position oscillation period is twice the power oscillation period, which is equal to the oscillation period of the light polarization.

The negligible beam position shift in the double-pass configuration provides a new opportunity to deliver optical power via optical fiber.  
We compare the laser beam power of free-space incident $P_{\rm in}$ and single mode fiber (SM-980-5.8-125, Thorlabs) output $P_{\rm out}$ under single-pass and double-pass configurations and plot the results in Fig. 7(a) and (b). 
In a single-pass configuration, the output beam power drops significantly as the quartz angle increases, contrasting with the result of the double-pass configuration. 
The high stability of the beam power in the double-pass configuration is shown in Fig.~7(c), where the coupling efficiency, $P_{\rm out}/P_{\rm in}$, remains almost constant in the range of (0.65, 0.66) [Fig.~7(c)].

\begin{figure}[htbp]
\centering\includegraphics[width=\linewidth]{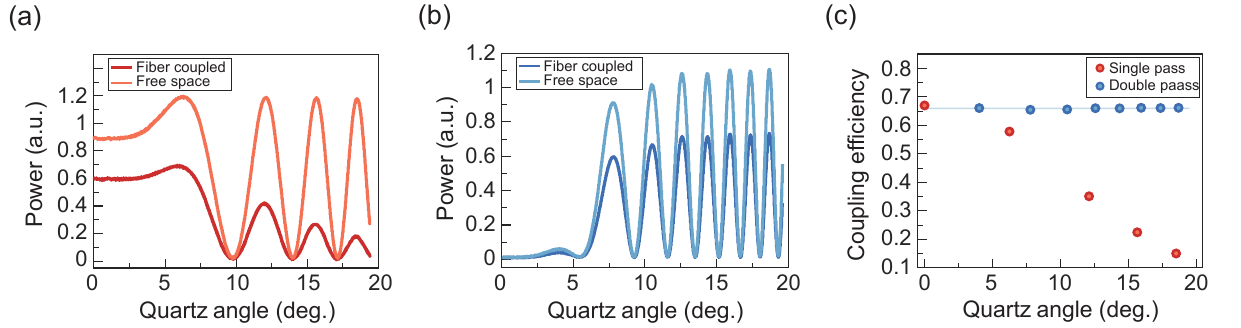}\label{Fig7}
\caption{(a) Comparison of free space incident power and fiber coupled power of single-pass configuration(S-setup). Measured power curves are obtained by averaging 10 measurements. 
(b) Comparison of free space incident power and fiber-coupled power of double-pass configuration(D-setup). Measured power curves are obtained by averaging 20 measurements. 
Power curves deviate from Eq.(1) when the quartz angle is lower than approximately $10^\circ$. We discuss this issue in detail in \textit{Appendix II}. 
(c) Fiber coupling efficiencies at power maxima points of single- and double-pass configurations. The blue shaded line represents the mean value and standard deviation of the coupling efficiency in the double-pass configuration.}
\end{figure}

\section{Rotating Z-cut quartz as rapidly variable waveplate}\label{Sec4}
The above discussions show that a rotating z-cut quartz plate functions as an effective variable waveplate. 
Noting that a half-wave retardation requires a rotation of less than $1^\circ$ in the double-pass configuration [Fig.4(b)], we consider the optical setup can be used as a rapid power converting system. 
In this section, we present the applicability of the optical setup for a rapid power converter.

\subsection{Z-cut quartz as a rapid power converter}
To test the rapid power conversion, we send a square-shaped trigger pulse to the galvanometer [Fig.~2(b)], which rotates the quartz crystal to have the $\pi/2$ retardation in the D-setup.
Specifically, it corresponds to an angular interval $\theta_f-\theta_i=0.7^\circ$ in Fig.~4(b). 
Upon the trigger pulse, we find the optical power reaches its full power within 1~ms [Fig.~8(a)]. 
Although the input square pulse causes vibration of the quartz plate when it reaches the stopping angle, it could be mitigated by shaping the trigger pulse. 
For instance, the sudden jump in the square pulse can be smoothed by adding an RC circuit with $\tau=RC=0.2$~ms, and the power conversion occurs within 2.5~ms. 
The galvanometer used to rotate the quartz can operate with input sine-wave signals at a rotation angle of $0.7^\circ$ and frequencies up to 300 Hz, which is limited by its mechanical nature. 

\begin{figure}[htbp]
\centering\includegraphics[width=\linewidth]{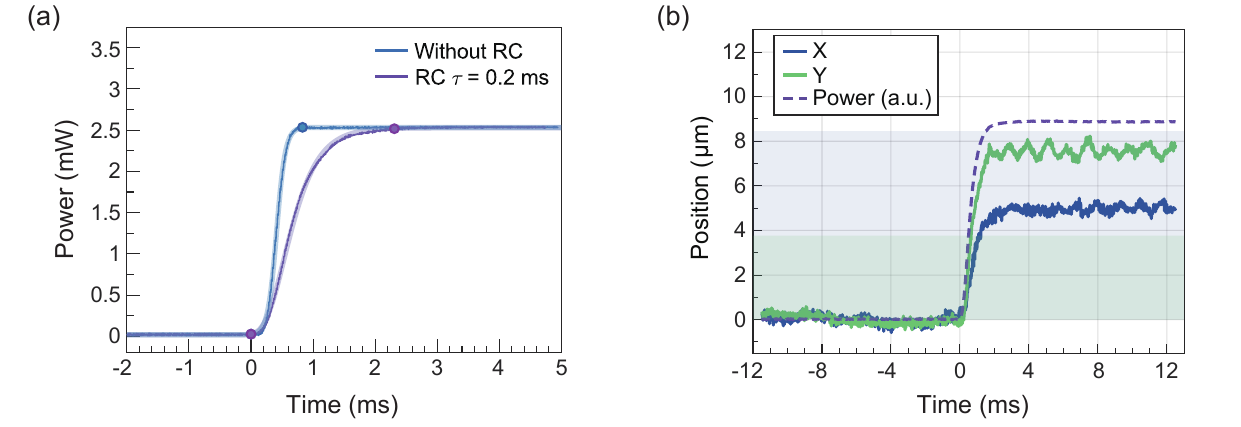}\label{Fig8}
\caption{(a) Power during the conversion using the D-setup, without (blue solid line) and with (purple solid line) an RC circuit. 
Measured power curves are obtained by averaging 10 measurements. 
The power curve without RC is fitted using the hyperbolic tangent function, $P(t) = P_0\tanh\left[(t - t_0)/\tau_{\rm{fit}}\right] + P_{\rm{off}}$, where $P_0 = P_{\mathrm{max}}/2$. In contrast, the power curve with the RC circuit fits well to a Gompertz function, $P(t) = P_{\rm{max}} \exp\left[-t_0 \exp(-t/\tau_{\rm{fit}})\right] + P_{\rm{off}}$. These fitting functions are shown in the figure as shaded lines. 
The power rise time, defined as the time required to increase from 0.3\% to 99.7\% of $P_{\rm{max}}$ based on the fitted function, increased from 0.83 ms to 2.31 ms after implementing the RC regulation circuit. These durations are indicated by the markers in the figure.
(b) Beam shift during the power conversion. To obtain clear position signals, an RC circuit is used. 
The shading represents the beam shift caused by the ordinary HWP. 
Measured power curves are obtained by averaging 30 measurements and applying a moving average filter with a window size of 5 to reduce background electronic noise.
}
\end{figure}

We characterize the power conversion of the optical system by measuring the ratio of beam power between two polarizations after the PBS in the D-setup. 
The conversion ratio is determined as $R_P \equiv (P_{\mathrm{max}} -P_{\mathrm{bg}})/(P_{\mathrm{min}}-P_{\mathrm{bg}})$, where $P_{\mathrm{max}}(P_{\mathrm{min}})$ is the maximum (minimum) beam power after (before) the rotation, and $P_{\mathrm{bg}}$ is the background noise without the laser beam. 
We measure all of the beam powers for 10 repetitions and obtain the $R_P$, which yields $R_P\sim 1000$. 
We also find that the robustness of the optical system, such that the conversion ratio can be maintained $R_P>900$ even when the retro-mirrors are deviated $\pm 0.015^{\circ}$ from the optimal point. 
In our setup, it corresponds to a quarter turn of the retro-mirror mount (MDI-170 TPI, Radiant-dyes).

We also monitor the beam position shift during power conversion by picking up the retro-reflected beam with a beam splitter in the DBS setup.
For this purpose, the RC circuit is implemented to obtain a clearer position signal by reducing vibrations. 
The average position in X and Y shows a clear deviation before and after the rotation Fig.~8(b), and the magnitude of the displacement is $9.1~\mu{}$m.
Considering the beam position shift from the beam splitter, we expect the positional shift from the rotating quartz to be less than 9.3~$\mu$m, measured using a zero-order HWP.
This is negligible beam shift compared to the $1/e^2$ beam diameter of 2.2~mm.

\begin{figure}[t]
\centering\includegraphics[width=0.95\linewidth]{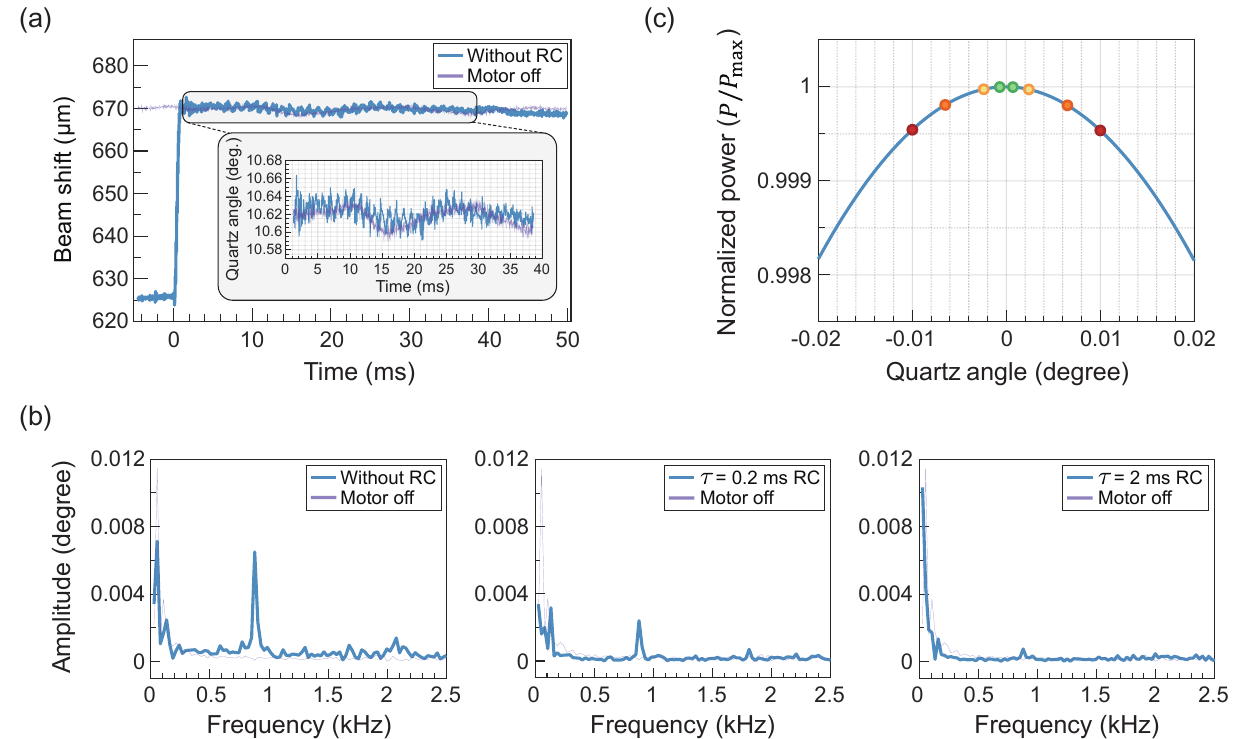}\label{Fig9}
\caption{(a) Measured single-passed beam shift to evaluate the vibration of the quartz angle. Input square signal rotates the quartz by $0.7^\circ$. 
The purple line represents the background beam position fluctuations, which is measured by turning off the motor.
The inset is converted beam shift into the quartz angle using the linear relation $\theta= 15.51(D+0.0149)$ where $D$ and $\theta$ represent the beam shift and the quartz angle with units of [mm] and [degree], respectively. 
(b) Frequency spectrum of quartz vibration for a bare square pulse, a rounded square pulse with $\tau = 0.2\,\mathrm{ms}$, and a $\tau = 2\,\mathrm{ms}$. Data are obtained by averaging 30 measurements. 
(c) Enlarged theoretical power maximum point near $\theta=16^\circ$ and markers to estimate the power fluctuation after the power conversion. Green, yellow, orange, and red markers are marked at $\pm\,0.00074^\circ$, $\pm\,0.0024^\circ$, $\pm\,0.0064^\circ$, and $\pm\,0.01^\circ$ of quartz angle, respectively.
}
\end{figure}
 
\subsection{Power fluctuations after a rapid rotation}
A rapid motional change in the galvanometer can cause angular vibration of quartz, leading to beam power fluctuations. 
To quantify the power fluctuations caused by the rotation, we study the fluctuations in the beam position shift at the stopping angle in a single-pass configuration (S-setup). 
It comes from the observation that the position shift in the S-setup can be directly matched to the rotational angle Fig.~6(b). 
Compared to directly measuring the beam power, this method has a much better precision in estimating the power fluctuations.
This is because the angle for the $\pi/2$ retardation is an extreme point in the power curve so that it is insensitive to the angular vibration.
Moreover, the intensity fluctuations are turned out to be small (less than $0.1\%$), and thus it is very challenging to distinguish the power fluctuations from the angular vibrations of the motor and the feedback signals from the intensity stabilization servo. The beam point is measured by the 2D lateral sensor, which is installed at the PD position of the S-setup. Fig.~9(a) displays the beam position shift as a function of time after the square trigger pulse. The displacement rapidly increases after the trigger and reaches a steady value within 1~ms. 
In the steady value, we find a clear oscillation in the beam position, implying the vibration of the motor from the rapid motion. 
The gradual change in the steady value represents the beam point drift in our optical setup, which is also shown even without the motor, Fig.~9(a). 

To characterize the power fluctuations, we investigate the frequency spectrum of the angular fluctuations of the quartz at stopping angle with different input signals [Fig.~9(b)]: a bare square pulse, smoothed pulses with $\tau = 0.2\,\mathrm{ms}$, and a $\tau = 2\,\mathrm{ms}$. 
Compared to the background noise, we find the two dominant frequencies around 1 and 2 kHz, but those peaks are substantially decreased with the regulated pulse. 
Markers in Fig.~9(c) represent the power range corresponding to the maximum angle amplitude near 1 kHz for each input signal. 
Note that the power curve is insensitive to the angular variations, where angular fluctuations $\Delta \theta =\pm 0.02^{\circ}$ correspond to 0.2\% in relative power fluctuations.
In our experiments, the estimated power fluctuation is less than 0.1\% of $P_{\mathrm{max}}$ for all input signals.
This result indicates that the noise level of the optical system is low enough to be applicable in an atomic quantum simulating platform. 
For example, it can be utilized to perform an early stage of evaporation cooling in an optical dipole trap~\cite{Burchianti2014}, to freeze out atomic motions in an optical lattice~\cite{Kwon2022} and to change the lattice geometry by adjusting the laser beam power of retro-reflected light~\cite{Tarruell2012}.

\section{Conclusion}
We have demonstrated a rapid control of light polarization by using a rotating z-cut quartz plate in a double-pass configuration. 
The variation of the polarization can be understood in the Jones matrix formalism, where we verify it through measurement of the beam power transmitted through a polarizing beam splitter as a function of the quartz angle.
Compared to conventional single-pass setups, the double-pass configuration significantly reduces beam displacement, down to below $10~\mu{}m$, which is sufficient for fiber coupling and high-stability applications.
In combination with a polarizing beam splitter, the optical system behaves as a compact and robust power modulator, reaching its full power within 1~ms with a dynamic range of 1000:1. 
We find the power fluctuations caused by the rapid motion, which is 0.1$\%$ of the maximal power, and it can be further mitigated by increasing the rise time. 
The optical system can be used for a wide wavelength range, high-power circumstances, and large-size incident beams, such as those with a diameter of about 20 mm in our case. This could be an option for situations requiring large beam sizes that cannot be addressed by an acousto-optic modulator. 
These features make the device particularly suitable for AMO physics applications that require fast, precise, and broadband polarization and intensity control.
\begin{backmatter}

\bmsection{Funding}
National Research Foundation of Korea (NRF) Grant under Projects No. RS-2023-00207974, RS-2023-00218998, RS-2023-00256050, and 2023M3K5A1094812. 

\bmsection{Acknowledgment}
The authors thank Junhyeok Hur and Deok-Young Lee for helpful discussions and technical contributions. We also appreciate Dr. Shribak for his valuable advice regarding optical activity.

\bmsection{Disclosures} The authors declare no conflicts of interest.

\bmsection{Data Availability Statement} Data underlying the results presented in this paper are not publicly available at this time but may be obtained from the authors upon reasonable request.
\end{backmatter}


\section*{Appendix I. Alignment process to use the z-cut quartz plate}\label{App1}
In this appendix, we introduce the procedure for aligning the double-passed beam.
As a first step, single-pass alignment should be achieved. As mentioned in section 2, the retardation $\Delta \phi(\theta)$ is symmetric~\cite{Paranin2016} for $\theta$, which satisfies $\Delta \phi(\theta)=\Delta \phi(-\theta)$. To utilize this property, the motor is fixed so that the full quartz angle sweep range ($\sim20^\circ$) encompasses the first and its pair power extrema, as shown in Fig.~10(a). The incident beam is aligned to make the two included power peaks symmetric by modifying the kinematic mount in the S-setup. The maximum of the power curve shown in Fig.~10(a) remains stable during this process, due to the robustness of the PBS cube against slight deviations from the optimal angle of incidence. From the well-aligned symmetric power curve, we can identify the two maxima [green dots in Fig.~10(a)] and determine the center point (blue dot), defined as the midpoint of the two maxima. We define the center point as $\theta=0^\circ$ from the symmetry, and use the angle interval $\Delta \theta$ between the first maxima and center point as the angle reference.

\begin{figure}[htbp]
\centering\includegraphics[width=\linewidth]{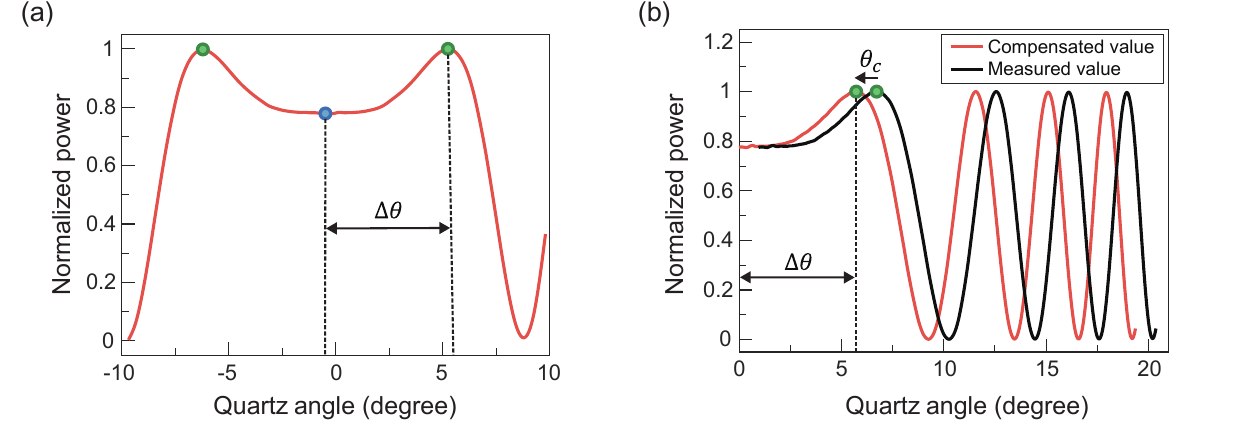}\label{Fig10}
\caption{
(a) Symmetric power curve to obtain the angle reference $\Delta \theta$ defined as an angle interval between first maxima and center point. (b) To make the motor angle to the quartz angle, angle reference $\Delta \theta$ is used. First maxima of the measured power curve is shifted to the $\Delta \theta$, and the shifted angle is named as the compensation angle $\theta_c$. In this figure, compensation angle is set by $\theta_c = 1^\circ$ for visibility. For the actual measurement in the main contexts is $\theta_c = 0.0475^\circ$. Power curves are obtained by averaging 10 measurements and applying a smoothing spline to find the maxima. 
}
\end{figure}

After finding the reference angle, we adjust the motor to approximately cover the angle range from zero to about $20^\circ$. Typically the motor angle is not exactly the same as the quartz angle.
Therefore, we need to find the offset angle of the motor to match the quartz angle $\theta$. 
For example, when the motor angle signal is 0 V, the quartz angle could be $1^\circ$ instead of $0^\circ$. 
To match the motor angle to quartz angle, power curve is shifted so that the first maxima point is at the $\Delta \theta$ as shown in Fig.~10(b). From this process, we obtain the compensated angle $\theta_c$, which is the deviation between the measured first maxima point and $\Delta \theta$. With this $\theta_c$, the motor angle signal can correspond to the quartz angle until the motor is modified. 
By slightly adjusting the kinematic mount to align the measured power curve with the theoretical curve as much as possible, the final single-pass power curve is obtained, as shown in Fig.~4(a).  For the double-pass configuration, alignment is achieved by adjusting the retro-mirror in the D-setup while keeping the PBS mount fixed. The retro-mirror is aligned so that every power extremum peak is as consistent as possible and aligns with the theoretical curve. To ensure proper alignment, we verify that the incident and reflected beams overlap using an IR viewer. After this alignment, the power curve is obtained as shown in Fig.~4(b).

\section*{Appendix II. Discrepancy at small quartz angle: Optical rotatory power}\label{App2}
\begin{figure}[htbp]
\centering\includegraphics[width=0.95\linewidth]{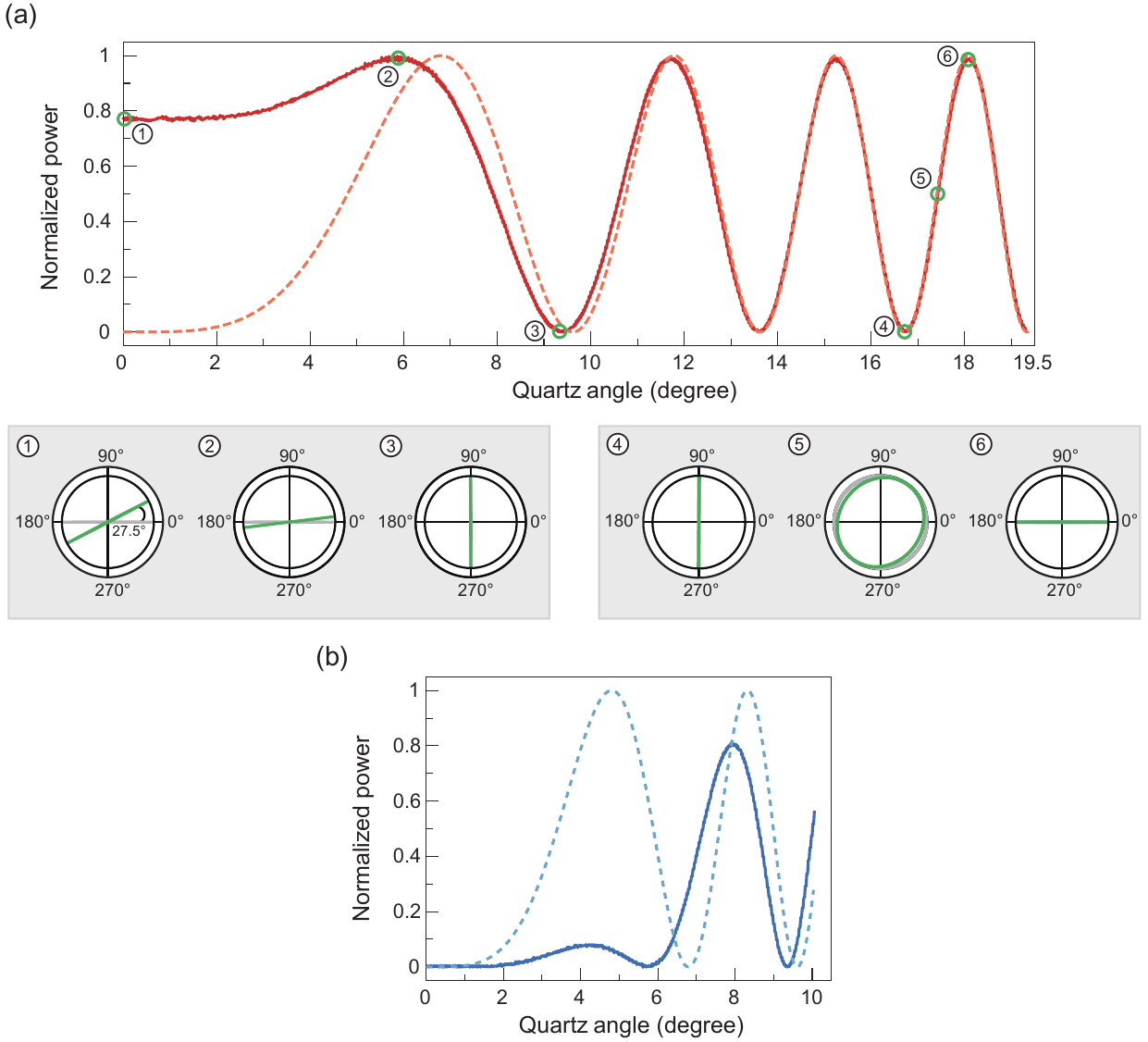}\label{Fig11}
\caption{
Beam power curves for the (a) single-pass and (b) double-pass configuration. 
Solid lines represent the measured power and dashed line is the theoretical curve without considering the optical rotation effect.
Insets in (a) are the polarization of the light at extreme points of the power curve.
Polarizations are measured by using a polarimeter (PAX1000IR1, Thorlabs) and are averaged over 60~s.
}
\end{figure}

As shown in Fig.~7, we find a discrepancy between the measured beam power and the theoretical expectation from Eq.(1) in the quartz angle range of 0° to 10°. 
It can also be seen in Fig.~11, where we plot power curves for the single- and double-pass configurations.
Moreover, the light polarization at local extremum angles of beam power is largely different from the expected polarizations. 
This discrepancy, observed for quartz angle below $10^\circ$, arises from optical rotatory power (ORP), which rotates the polarization plane of incident light~\cite{Condon1937}. 
The z-cut quartz is optically active and has been widely used for studying ORP~\cite{Vedam1958,Vido2019,Arteaga2009} since the birefringence effects can be eliminated by sending a light with a normal angle.
Our z-cut quartz has an ORP of $6.3^\circ/\mathrm{mm}$ at room temperature, resulting in a total polarization rotation of $63^\circ$. 
It implies the incident S-polarization is rotated by $63^\circ$ even when the quartz angle is zero. 
In other words, the angle between the rotated polarization of light and the P-polarization becomes $27^\circ$. 
As shown \circled{1} in Fig.~11(a), the measured angle between them is $27.5(2)^\circ$, leading us to conclude that this discrepancy arises from the ORP. The ORP decreases as the wavelength and incident angle of light increase, as reported in a previous study\cite{Jazi2019}. For this reason, in the quartz angle range beyond $10^\circ$ as described in the main text, the birefringence effect dominates over the ORP. Consequently, the measured power curve fits well with Eq. (1), which is derived solely from the birefringence effect. 

In the single-pass configuration, linear polarization should be converted to circular polarization from the retardation. 
In the range of quartz angle over $10^{\circ}$, polarizations at principal points are measured and the results are shown in Fig.~11(a). 
The conversion between P-polarization and S-polarization is achieved similarly to a zero-order waveplate. 
However, the conversion from linear to circular polarization is not fully achieved due to the optical activity of quartz. 
The maximum ellipticity of the polarization at point \circled{5} in Fig.~11(a) is measured to be $41.7^{\circ}$, which deviates from $45^{\circ}$, the ideal value for circular polarization. 
This conversion requires optimizing the quartz thickness in consideration of its optical activity\cite{Shribak1986}. \\
The deviation between Eq.(1) and the measured power curve also appears in the double-pass configuration, as shown in Fig.~11(b). 
This suggests that materials other than quartz may be preferable to have a better control of beam polarization. 
That is, when thermal lensing and cost are not critical considerations for the intended application, other materials such as $\mathrm{MgF}_2$ or sapphire may be more suitable. 
These materials exhibit no optical activity, and thus the power curves could follow Eq.(1) more closely, particularly at rotation angles below $10^\circ$. Materials with larger birefringence ($\Delta n = n_o - n_e$), such as $\alpha$-BBO and $\mathrm{YVO}_4$, can also be considered. These allow power conversion with rotation angles smaller than $0.7^\circ$. 
However, such small rotation angles may result in increased sensitivity to angular fluctuations. 
Therefore, the choice of material for the power conversion must be made based on the specific requirements and constraints of the intended application.

\bibliography{Ref_Quartz}

\end{document}